\documentclass[twocolumn]{aastex631}

\usepackage{xcolor}
\usepackage{amsmath}

\newcommand{\muygpys}{\texttt{MuyGPyS}}
\newcommand{\treecorr}{\texttt{TreeCorr} }

\begin{document}

\title{A Scalable Gaussian Process Approach to Shear Mapping with MuyGPs}

\author{Gregory Sallaberry}
\affiliation{Lawrence Livermore National Laboratory \\
	7000 East Avenue \\
	Livermore, CA 94550, USA}

\author{Benjamin W. Priest}
\affiliation{Lawrence Livermore National Laboratory \\
	7000 East Avenue \\
	Livermore, CA 94550, USA}

\author{Robert Armstrong}
\affiliation{Lawrence Livermore National Laboratory \\
	7000 East Avenue \\
	Livermore, CA 94550, USA}

\author{Michael Schneider}
\affiliation{Lawrence Livermore National Laboratory \\
	7000 East Avenue \\
	Livermore, CA 94550, USA}

\author{Amanda Muyskens}
\affiliation{Lawrence Livermore National Laboratory \\
	7000 East Avenue \\
	Livermore, CA 94550, USA}

\author{Trevor Steil}
\affiliation{Lawrence Livermore National Laboratory \\
	7000 East Avenue \\
	Livermore, CA 94550, USA}

\author{Keita Iwabuchi}
\affiliation{Lawrence Livermore National Laboratory \\
	7000 East Avenue \\
	Livermore, CA 94550, USA}
	
\begin{abstract}
	Analysis of cosmic shear is an integral part of understanding structure growth across cosmic time, which in-turn provides us with information about the nature of dark energy.
	Conventional methods generate \emph{shear maps} from which we can infer the matter distribution in the universe.
	Current methods (e.g., Kaiser-Squires inversion) for generating these maps, however, are tricky to implement and can introduce bias.
	Recent alternatives construct a spatial process prior for the lensing potential, which allows for inference of the convergence and shear parameters given lensing shear measurements.
	Realizing these spatial processes, however, scales cubically in the number of observations - an unacceptable expense as near-term surveys expect billions of correlated measurements.
	Therefore, we present a linearly-scaling shear map construction alternative using a scalable Gaussian Process (GP) prior called MuyGPs.
	MuyGPs avoids cubic scaling by conditioning interpolation on only nearest-neighbors and fits hyperparameters using batched leave-one-out cross validation.
	We use a suite of ray-tracing results from N-body simulations to demonstrate that our method can accurately interpolate shear maps, as well as recover the two-point and higher order correlations.
	We also show that we can perform these operations at the scale of billions of galaxies on high performance computing platforms.
\end{abstract}



\section{Introduction}\label{sec:intro}
Weak lensing is a powerful tool for understanding the matter distribution in the universe.
Cosmic shear, the distortion of the true ellipticity of background sources by intermediate large scale structure, allows us to probe the growth of structure across cosmic time.
Tracking this structure growth gives insight into the temporal evolution and nature of dark energy and can help provide important constraints on cosmological parameters.
These parameters include the cosmic matter and dark energy densities $\Omega_m$ and $\Omega_{\Lambda}$, respectively, and the linear spectrum of mass fluctuations $\sigma_8$.

Unlike probes such as the cosmic microwave background (CMB), weak lensing benefits from information on much smaller angular scales, i.e., scales in the nonlinear regime of structure formation.
Weak lensing allows the reconstruction of the cosmic matter distribution across a wider distribution of redshifts, including in regions at $z<1.0$.
This mass reconstruction is agnostic to assumptions about the relationship between baryonic and dark matter and the nature of dark matter~\citep{schneider2006introduction}.

Our objective while studying weak lensing is the construction of a \emph{shear map} - a mapping of unlensed to lensed ellipticity as a function of sky coordinates.
Notationally, we consider shear maps to be representations of the two-dimensional lensing distortions along the pure $(x,y)$ directions, $\gamma_1$, and along the diagonals, $\gamma_2$.
Complimentary to the maps of shear, we have convergence $\kappa$, which refers to the two-dimensional surface mass density corresponding to the weak lensing distortions $\gamma_1$ and $\gamma_2$.
Where applicable, the combination of the shear and convergence maps will be referenced collectively as simply the full set of \emph{shear maps}.

A shear map allows us to infer the \emph{lensing potential} $\psi$ of points in the night sky, by way of its relation to $\kappa$, $\gamma_1$, and $\gamma_2$ through a combination of second-order partial derivatives (the lensing equations, Equations~\eqref{eq:potential_odes}).
The relationship between these quantities can be exploited to use data from shear measurements to reconstruct the two-dimensional mass density in a given region of space.

This mass reconstruction, however, is a fraught procedure.
For any sample of background galaxies, there is uncertainty in the intrinsic shape of the sources, the so-called \emph{shape noise}, as well as in other cosmological parameters.
Additionally, the \emph{method} of mass reconstruction can introduce systematics.
For example,~\cite{Kaiser_1993ApJ} introduced a method of inverting the shear field to recover the convergence, but the method struggles in situations with a finite survey area or incomplete coverage (e.g., masks).

Recent studies have used probabilistic frameworks to generate shear maps in a systematic-free way, with the aim of capturing the non-gaussianity of the cosmic matter field at small scales. 
In the non-gaussian (nonlinear) regime, higher order statistical information has been demonstrated to be useful in providing constraints on various cosmological parameters~\citep{gatti_2020MNRAS}.
Such methods have included using field-level inference using log-normal priors~\citep{zhou_2024PhRvD} and a bayesian hierarchical approach with similar priors~\citep{fiedorowicz_2022MNRAS, boruah_2024PhRvD}.

~\cite{Schneider_2017ApJ} proposed a Gaussian Process (GP) prior for the lensing potential and demonstrated the ability to generate convergence maps in the limit of low galaxy shape noise.
They presented a GP kernel motivated by the relationship between shear and convergence in the lensing equations.

However, next-generation surveys such as the Legacy Survey of Space and Time (LSST) to be performed at the Vera C. Rubin Observatory will capture shape information of billions of correlated sources across the survey lifetime.
Conventional GP interpolation, though valued for its flexibility, expressiveness, and natural uncertainty quantification, requires quadratic memory and cubic computation.
Thus, a straightforward application of the method described in~\cite{Schneider_2017ApJ} will not scale to next-generation survey data sizes.

The purpose of this work is to demonstrate a scalable alternative spatial process prior for correlated lensing shear parameters that can be the basis of future investigations.
We make several simplifying assumptions on the statistical and astronomical models throughout in order to achieve this objective with maximum clarity.
This framework can be utilized in subsequent work to examine to the cosmic matter distrubution on the largest of available scales in the effort of further constraining the values of $\Lambda CDM$ cosmological parameters.

In order to achieve billion-scale computation of shear maps, we employ an alternative GP model called ``MuyGPs''.
\cite{muyskens2021muygps} shows that MuyGPs scales linearly in the amount of data by inducing nearest neighbors-based sparsity in the correlation function of the GP.
We also utilize state-of-the-art distributed memory nearest neighbors algorithms to allow our model realizations to scale to modern high performance computing systems.

We organize the work as follows.
Section~\ref{sec:method} introduces our statistical and computational models.
Section~\ref{sec:data} discusses our analytical framework and justifies our model decisions with validation tests.
Finally, we present our scaling and experimental results in Section~\ref{sec:results} before our final remarks in Sections~\ref{sec:summary} \&~\ref{sec:future}.


\section{Method}\label{sec:method}

We first introduce the spatial process prior used by~\cite{Schneider_2017ApJ}.
We then illustrate how our method improves upon its scalability by employing MuyGPs and discuss implementation details on high performance computing platforms.

\subsection{A Statistical Model of Lensing Shear}

Most of the lensing distortion of source ellipticity described in Section~\ref{sec:intro} falls into the regime of \emph{weak lensing}.
Contrary to \emph{strong lensing}, where the distortions of point sources are clearly detectable, sometimes even by the human eye, the \emph{weak lensing} phenomena is only measurable in the form of correlated alignment present in the ellipticity measurements of neighboring galaxies - i.e., the measurement of weak lensing is inherently statistical.

We will posit the following spatial process prior for our shear maps, following~\cite{Schneider_2017ApJ}.
We assume that each of $n$ galaxies are imaged, with measured point spread functions and Gaussian shape noise.

We will endeavor to map lensing shear at a sky coordinate $\mathbf{x} = [x_1, x_2]$.
A shear map at $\mathbf{x}$ aims to recover three parameters: the convergence $\kappa(\mathbf{x})$ and the two orthogonal shear parameters $\gamma_1(\mathbf{x})$ and $\gamma_2(\mathbf{x})$.
For convenience, we write the shear measurements $\mathbf{x}$ as $\Upsilon(\mathbf{x}) = [\kappa(\mathbf{x}), \gamma_1(\mathbf{x}), \gamma_2(\mathbf{x})]$.
Throughout this paper we will drop the dependency on $x$ where it is clear from context.
We also make some simplifying assumptions for the purpose of this analysis:
\begin{enumerate}
	\item intrinsic (unlensed) ellipticity and measurement error for all galaxy observations are drawn from a prior Gaussian distribution with prior variance $\sigma_g$.
	\item The measurement locations $\mathbf{x}$ are recorded on a 2-dimensional plane, rather than using spherical coordinates\footnote{It is straightforward, to generalize the analysis to spherical coordinates, but for simplicity we assume we can describe our coorinates in flat Minkowski space.}.
	\item Galaxy observations include measurements of $\kappa$, $\gamma_1$, and $\gamma_2$.
	\item The noise model is homoscedastic Gaussian with prior variances $\sigma_\kappa$, $\sigma_{\gamma_1}$, and $\sigma_{\gamma_2}$, respectively.
\end{enumerate}
The purpose of these assumptions is to simplify the presentation and analysis, as the primary contribution of this work is the scalability of the shear map realization.
We will discuss future generalizations to our analysis that divest these assumptions in Section~\ref{sec:summary}.

As discussed above, measurements of shear at location $\mathbf{x}$, $\kappa$, $\gamma_1$, and $\gamma_2$ are in fact proxies for the \emph{lensing potential} $\psi(\mathbf{x})$ - our quantity of interest.
The lensing potential $\psi$ is related to each of the shear parameters $\kappa$, $\gamma_1$, and $\gamma_2$ by the following equations:
\begin{align}
	\begin{split}
		\kappa &= \frac{1}{2} \left (
			\frac{\partial^2\psi}{\partial^2 \mathbf{x}_1}
			+ \frac{\partial^2\psi}{\partial^2 \mathbf{x}_2}
		\right ), \\
		\gamma_1 &= \frac{1}{2} \left (
			\frac{\partial^2\psi}{\partial^2 \mathbf{x}_1}
			- \frac{\partial^2\psi}{\partial^2 \mathbf{x}_2}
		\right ), \\
		\gamma_2 &= \frac{1}{2} \left (
			\frac{\partial^2\psi}{\partial \mathbf{x}_1 \partial \mathbf{x}_2}
			+ \frac{\partial^2\psi}{\partial \mathbf{x}_1 \partial \mathbf{x}_2}
		\right ),
	\end{split} \label{eq:potential_odes}
\end{align}
herein referred to as the \emph{lensing equations}.
For proper treatment and derivation of these equations under the Born Approximation~\citep{born2013principles}, see~\cite{schneider2006introduction}.
It is straightforward to show using Equations~\eqref{eq:potential_odes} that a GP prior on $\psi$ implies a GP prior on $\Upsilon$.
We must now introduce some Gaussian process terminology to proceed.

A Gaussian process is a parameterized distribution over functions $\mathcal{GP}(\mu(\cdot), k_\theta(\cdot, \cdot))$, where the mean function $\mu \equiv \mathbf{0}$ without a loss a generality and $k_\theta(\cdot, \cdot)$ is a positive-definite covariance function that is parametrically controlled by hyperparameters $\theta$.
In our setting, the lensing potential map $\psi$ is GP distributed if any finite set of sky coordinates $X = [\mathbf{x}^{1}, \dots, \mathbf{x}^{n}]$ has corresponding lensing potentials
\begin{equation} \label{eq:gp_prior}
\psi(X) = \left (
	\psi(\mathbf{x}^{1}), \dots, \psi(\mathbf{x}^{n})
\right )^\top
\sim \mathcal{N} \left (\mathbf{0}, k^\psi_\theta(X, X) \right ).
\end{equation}
Here $\mathcal{N}$ is the multivariate Gaussian distribution, and $k^\psi_\theta(X,X)$ is a square covariance matrix whose $i,j$th element is
$k^\psi_\theta(\mathbf{x}^{(i)},\mathbf{x}^{(j)})$ for the kernel function $k^\psi_\theta$.

We follow the treatment in~\cite{Schneider_2017ApJ} and impose a squared exponential or radial basis function (RBF) form for the kernel function:
\begin{equation} \label{eq:kernel_psi}
	k^\psi_{\lambda, \ell}(\mathbf{x}, \mathbf{y})
	= \lambda^{-1} \exp \left(
		-\frac{s^2(\mathbf{x},\mathbf{y})}{2\ell^2}
	\right),
\end{equation}
where $\lambda$ is a variance scaling parameter, $\ell$ is a length scale parameter, and
$s^2(\mathbf{x}, \mathbf{y}) \equiv (\mathbf{x} - \mathbf{y})^\top(\mathbf{x} - \mathbf{y})$.
\cite{Schneider_2017ApJ} chose the RBF because it is an especially smooth kernel that is applicable to low-resolution or low signal-to-noise ratio observations, which we expect to apply to our demonstrations as well as future applications.
Practically, the RBF is also useful for its parametric simplicity and the relatively straightforward form it implies for the kernel function of $\Upsilon$ based upon Equations~\eqref{eq:potential_odes}.

The GP prior on $\psi$ implies a GP prior on $\Upsilon$ of the form
$\mathcal{GP} \left (\mathbf{0}, k^\Upsilon_{\lambda, \ell}(\cdot, \cdot) \right )$.
$k^\Upsilon$ has the following form (where we have dropped the $\lambda$, $\ell$ subscripts and $\mathbf{x}$, $\mathbf{y}$ arguments for clarity):
\begin{equation} \label{eq:kernel_upsilon}
k^\Upsilon = \left(\begin{array}{ccc}
  k^{\kappa, \kappa}    &  k^{\kappa, \gamma_1}    &  k^{\kappa, \gamma_2} \\
  k^{\gamma_1, \kappa}  &  k^{\gamma_1, \gamma_1}  &  k^{\gamma_1, \gamma_2} \\
  k^{\gamma_2, \kappa}  &  k^{\gamma_2, \gamma_1}  &  k^{\gamma_2, \gamma_2} \\
\end{array}\right),
\end{equation}
where $k^{\alpha, \beta} = k^{\beta, \alpha}$ is the covariance function associated with covariates $\alpha$ and $\beta$.
The form of the functions in Equation~\eqref{eq:kernel_upsilon} are implied by
combining Equation~\eqref{eq:kernel_psi} with Equations~\eqref{eq:potential_odes}.
The full derivations can be found in~\cite{NG_2016PhDT} and restated in Appendix A of~\cite{Schneider_2017ApJ}

Finally, we can state the full joint covariance of the shear parameters at two
sky locations, $\mathbf{x}$ and $\mathbf{y}$:
\begin{align}
	\begin{split}
	\Pr (
		\Upsilon(\mathbf{x}), \Upsilon(\mathbf{y})
		&\mid \mathbf{d}, \sigma_g, \lambda, \ell
	) \\
	=& \mathcal{N}_{\Upsilon(\mathbf{x}), \Upsilon(\mathbf{y})} \left (
		\mu^\Upsilon, \Sigma^\Upsilon
	\right ) \\
	&\times
	\mathcal{N}_\mathbf{\hat{e}} \left (
		\mathbf{0}, k^\Upsilon(X, X) + N
	\right ).
	\end{split} \label{eq:joint_distribution}
\end{align}
Here $\mathbf{d}$ represents all the observed points contributing to the measurements $\Upsilon(\mathbf{x})$ and $\Upsilon(\mathbf{y})$,
$\sigma_g$ is the prior variance of the Gaussian intrinsic (unlensed) galaxy shape distribution,
and $\mathbf{\hat{e}}$ is the mean of the observed (lensed) galaxy shapes.
Thus, $\mathbf{\hat{e}}$ is a $3n$ vector such that $\mathbf{\hat{e}} = [\kappa(x), \gamma_1(X), \gamma_2(X)]^\top$, whose values are observed.\footnote{In practice $\gamma_1$ and $\gamma_2$ are most easily measured while $\kappa$ is somewhat less straightforward to infer directly. Interpolating $\kappa$ from only shear requires a different form of $\mathbf{\hat{e}}$ and therefore Equations~\eqref{eq:kernel_upsilon} and~\eqref{eq:noise_prior}. We discuss a path to generalizing our methods in Section~\ref{sec:summary}.}
We use a noise prior covariance matrix $N$ of the form
\begin{equation} \label{eq:noise_prior}
	N = \left(\begin{array}{ccc}
		\sigma_\kappa * I_n & \mathbf{0}    &  \mathbf{0} \\
		\mathbf{0}  &  \sigma_{\gamma_1} * I_n  &  \mathbf{0} \\
		\mathbf{0}  &  \mathbf{0}  &  \sigma_{\gamma_2} * I_n \\
	\end{array}\right),
\end{equation}
where $\sigma_\kappa$, $\sigma_{\gamma_1}$, and $\sigma_{\gamma_2}$ are the homoscedastic prior variances of the measurements of each shear covariate.\footnote{This noise model can be make more general, as we discuss in Section~\ref{sec:summary}.}
Finally, we define $\mu^\Upsilon$ and $\Sigma^\Upsilon$ as the Gaussian process likelihood mean and variance given the GP prior for $\Upsilon$:
\begin{align}
	\mu^\Upsilon =
	k^\Upsilon(X, X) \left( k^\Upsilon(X, X) + N \right)^{-1} \mathbf{\hat{e}}
	\label{eq:mean_upsilon} \\
	\Sigma^\Upsilon =
	k^\Upsilon(X, X) \left( k^\Upsilon(X, X) + N \right)^{-1} N,
	\label{eq:var_upsilon}
\end{align}
where $k^\Upsilon(X, X)$ is a $3n \times 3n$ matrix that is structured blockwise like Equation~\eqref{eq:kernel_upsilon}.

Similar quantities can also be used to predict the shear parameters (and therefore the lensing potential) at unobserved sky coordinates $Z = (\mathbf{z}^{(1)}, \dots, \mathbf{z}^{(m)})$ via similar equations to obtain the posterior mean $\widehat{\mu}^\Upsilon = \mu^\Upsilon(Z \mid X)$ and posterior variance $\widehat{\Sigma}^\Upsilon = \Sigma^\Upsilon(Z \mid X)$:
\begin{align}
	\widehat{\mu}^\Upsilon &=
	k^\Upsilon(Z, X) \left( k^\Upsilon(X, X) + N \right)^{-1} \mathbf{\hat{e}}
	\label{eq:mean_upsilon_posterior} \\
	\widehat{\Sigma}^\Upsilon &=
	k^\Upsilon(Z, X) \left( k^\Upsilon(X, X) + N \right)^{-1} k^\Upsilon(X, Z),
	\label{eq:var_upsilon_posterior}
\end{align}
where $k^\Upsilon(Z, X) = k^\Upsilon(X, Z)^\top$ is an $3m \times 3n$ matrix that is structured blockwise following Equation~\eqref{eq:kernel_upsilon} similar to $k^\Upsilon(X, X)$.

\subsection{A Scalable Approximation using MuyGPs} \label{sec:muygps}

Unfortunately, computing Equations~\eqref{eq:mean_upsilon} and~\eqref{eq:var_upsilon} (required to sample from the joint distribution) or~\eqref{eq:mean_upsilon_posterior} and~\eqref{eq:var_upsilon_posterior} (required to interpolate to new coordinates) is intractable for large surveys due to their cubic complexity in $n$, the number of galaxy observations.
This computational bottleneck is a well-known shortcoming of Gaussian process interpolation that many investigators have attempted to overcome through model compromises of various types.
See, e.g., \cite{heaton2019case}, \cite{liu2020gaussian}, and \cite{muyskens2021muygps} for a survey of alternatives.
We dispense with a detailed comparison between methods and instead describe our chosen solution, MuyGPs.

MuyGPs is among a class of scalable GP models that impose sparsity in the GP covariance, meaning that it is not necessary to realize and invert $k^\Upsilon(X, X)$.
MuyGPs furthermore avoids evaluating a Gaussian likelihood to fit model hyperparameters $\lambda$ and $\ell$, which would require computing a determinant of $k^\Upsilon(X, X)$, in favor of optimizing a regularized loss function using batched leave-one-out cross-validation, see~\cite{muyskens2021muygps} and~\cite{wood2022scalable} for details of hyperparameter fitting.
MuyGPs has been applied to several uncertainty-sensitive applications at scales well beyond the reach of conventional Gaussian processes, including the modeling of space object light curves~\citep{goumiri2022light,goumiri2023light,goumiri2024uncertainty}, galaxy and star blend identification \citep{buchanan2022gaussian,eleh2024stellar}, star-galaxy image separation \citep{muyskens2022star}, atmospheric particle modeling \citep{mukangango2024robust} and electrocardiography anomaly detection \citep{nnyaba2024enhancing}, among others.

MuyGPs simplifies the GP posterior distributions of Equations~\eqref{eq:mean_upsilon_posterior} and~\eqref{eq:var_upsilon_posterior} by way of conditioning each point prediction only on its $p$ nearest neighbors in $X$, the set of observed galaxy sky coordinates.
This transforms the posterior equations for a single prediction location $\mathbf{z}$ to the following form:
\begin{align}
	\widehat{\mu}^\Upsilon(\mathbf{z} \mid X, p) &=
		C(\mathbf{z} \mid X, p) \mathbf{\hat{e}}_{J_{\mathbf{z}, p}}
	\label{eq:mean_posterior} \\
	\widehat{\Sigma}^\Upsilon(\mathbf{z} \mid X, p) &=
		C(\mathbf{z} \mid X, p) k^\Upsilon(X_{J_{\mathbf{z}, p}}, \mathbf{z}) \label{eq:var_posterior}
\end{align}
where
\begin{align}
	\begin{split}
		C(\mathbf{z} &\mid X, p) = \\
		&k^\Upsilon(\mathbf{z}, X_{J_{\mathbf{z}, p}})
		\left(
			k^\Upsilon(X_{J_{\mathbf{z}, p}}, X_{J_{\mathbf{z}, p}}) + N_{J_{\mathbf{z}, p}}
		\right)^{-1}.
	\end{split} \label{eq:kriging}
\end{align}
Here $J_{\mathbf{z}, p}$ is the set of indices of $\mathbf{z}$'s $p$ nearest neighbors in $X$.
Thus, $X_{J_{\mathbf{z}, p}}$ are the sky coordinates of those neighbors, $N_{J_{\mathbf{z}, p}}$ is a $3p \times 3p$ diagonal matrix that has been appropriately subselected from $N$, and similarly $\mathbf{\hat{e}}_{J_{\mathbf{z}, p}}$ is a $3p$ vector that has been appropriate subselected from $\mathbf{\hat{e}}$.
The power of MuyGPs comes from the observation that, for broad classes of practical problems, the \emph{kriging weights} defined as
\begin{align}
	C(\mathbf{z} &\mid X) =
	k^\Upsilon(\mathbf{z}, X)
	\left(
		k^\Upsilon(X, X) + N
	\right)^{-1} \label{eq:kriging_dense}
\end{align}
that occur in Equations~\eqref{eq:mean_upsilon}-\eqref{eq:var_upsilon_posterior} are \emph{sparse}.
That is, for stationary kernels and $X$ filling a large enough volume, there are only $q \ll n$ columns of the $3 \times 3n$ matrix $C(\mathbf{z} \mid X)$ defined in Equation~\eqref{eq:kriging_dense} with nontrivial norms.
This arises from the term $k^\Upsilon(\mathbf{z}, X)$, whose columns for all $\mathbf{x} \in X$ such that $s^2(\mathbf{z}, \mathbf{x})$ is large enough will be vanishingly small due to Equation~\eqref{eq:kernel_psi}.
This means that for large enough $p$, the nearest neighbor kriging weights $C(\mathbf{z} \mid X, p)$ are a good approximation of the true kriging weights, and can be used to interpolate as in Equations~\eqref{eq:mean_posterior} and~\eqref{eq:var_posterior}.

As a result of this sparsification, the posterior mean and variance can now be computed in $O(np^3)$ time instead of $O(n^3)$ time.
\cite{muyskens2021muygps} showed that, in practice, $p \ll n$ suffices to obtain high-quality predictions, which in turn implies that MuyGPs can scale to immense $n$.

\subsection{Implementation Details} \label{sec:implementation}

The scale of our demonstrations necessitates that our codes run on high performance computing (HPC) platforms in the distributed memory model.
In this model, our observation data $X$ is partitioned across a set of processors $\mathcal{P}$, which requires that the steps of computation take this into account so as not to waste resources on overlarge amounts of communication.
We will say that coordinate $\mathbf{x} \in X$ is assigned to processor $P_\mathbf{x} \in \mathcal{P}$ via a mechanism introduced below.

Our basic prediction workflow proceeds in three stages.
In stage one, a nearest neighbors index must be prepared using $X$ and a number-of-neighbors parameter $p$.
In stage two, this index is used to construct kernel tensors realizing the covariance $k^\Upsilon (X_{J_{\mathbf{z}, p}}, X_{J_{\mathbf{z}, p}})$ and cross-covariance $k^\Upsilon (\mathbf{z}, X_{J_{\mathbf{z}, p}})$, and the observed means $\hat{e}_{J_{\mathbf{z}, p}}$ and localizing them to $P_\mathbf{z}$ for all prediction sky coordinate $\mathbf{z} \in Z$.
In stage three, we compute Equations~\eqref{eq:mean_posterior} and~\eqref{eq:var_posterior} in parallel for each prediction coordinate $\mathbf{z}$.

Stage three is straightforward to perform in distributed memory once the crosscovariance and covariance tensors are localized at the end of stage two.
We used \muygpys, an open source implementation of the MuyGPs algorithm that is
intended for HPC distributions to perform these computations\footnote{The open source implementation of \muygpys is available at https://github.com/LLNL/MuyGPyS~\citep{priest2024muygpys}}.

However, constructing and querying the nearest neighbors index and constructing the tensors themselves (steps 1 and 2) are highly nontrivial.
While the fine details are out of the scope of this work, we will state the high-level algorithmic details here.
We used saltatlas, an HPC library implementing several scalable K-nearest neighbors algorithms in distributed memory~\citep{sanders2021saltatlas}\footnote{The open source implementation of saltatlas is available at https://github.com/LLNL/saltatlas~\citep{steil2024saltatlas}}.
We used saltatlas's implementation of DHNSW, saltatlas's distributed generalization of the popular approximate nearest neighbors algorithm Hierarchical Navigable Small Worlds (HNSW)~\citep{malkov2018efficient}.

HNSW is a state-of-the-art approximate nearest neighbor index that is favored in applications requiring fast queries and high recall.
HNSW operates by partitioning data into a logarithmic number of layers.
The bottom layer features all data points, while higher levels become logarithmically sparser.
Each layer constructs a nearest neighbor graph, and queries traverse these graphs from the top layer (sparse) to the bottom layer (fine) to return nearest neighbors using only a logarithmic number of comparisons rather than the worst case (linear).

DHNSW generalizes HNSW by creating a binary tree $T$ with a power-of-two set of seeds that iteratively partition the input space.
Each vertex of this tree imposes a metric hypersurface that iteratively partitions the input space.
Seeds are chosen to keep the resulting partitions nearly equal in size and to maintain locality among data points with relatively low joint metric distance in $s^2$.
Thus, the replicated tree assigns points $\mathbf{x} \in X$ to $T(\mathbf{x}) = P_\mathbf{x} \in \mathcal{P}$.
Each $P \in \mathcal{P}$ then maintains an HNSW data structure using the points in $X$ that are partitioned to $P$.
A query point $\mathbf{z}$ can be directed to a single HNSW instance $T(\mathbf{z}) = P_z \in \mathcal{P}$ by traversing the replicated tree, or the traversal can be truncated, sending $\mathbf{z}$ to several processors and HNSW instances and then consolidating the resulting neighborhoods into the closest $p$ points found.
This truncated traversal allows for high quality nearest neighbor queries for points that are near the edge of a partition at the expense of some additional communication between processors.
Additionally, positions of the points returned by queries against the HNSW instances can be used to suggest additional HNSW instances to query, exploring "neighbors of neighbors" as additional candidates.

Once DHNSW identifies a set of neighbor indices $J_{\mathbf{z}, p}$ for a query point $\mathbf{z}$, we identify $P_\mathbf{x}$ for each $\mathbf{x} \in X_{J_{\mathbf{z}, p}}$ and have it forward $\mathbf{x}$ and $\mathbf{\hat{e}}_\mathbf{x}$ to $P_\mathbf{z}$.
The implementation of DHNSW makes it simple to interleave hundreds of millions of these queries at once.
Once the neighborhood points $X_{J_{\mathbf{z}, p}}$, observations $\mathbf{\hat{e}}_{J_{\mathbf{z}, p}}$ and query point $\mathbf{z}$ are all colocated on the processor $P_\mathbf{z}$, it is simple to compute the covariance $k^\Upsilon (X_{J_{\mathbf{z}, p}}, X_{J_{\mathbf{z}, p}})$ and cross-covariance $k^\Upsilon (\mathbf{z}, X_{J_{\mathbf{z}, p}})$, followed by Equations~\ref{eq:mean_upsilon_posterior} and~\ref{eq:var_upsilon_posterior}, for all $\mathbf{z} \in Z$ in-place using MuyGPyS, completing the interpolation.

\section{Data and Validation}\label{sec:data}

For our analysis, we use data from the Scinet LIghtcone Simulations~\citep{Harnois-Deraps_2015MNRAS, Harnois-Deraps_2012MNRAS}: a set of ray tracing results based on a suite of N-body simulations, refered to hereafter as SLICS.
The underlying N-body simulations use a \textit{WMAP9} cosmology with $\Omega_m = 0.2905$, $\Omega_{\Lambda} = 0.7095$, $\Omega_b = 0.0473$, $h = 0.6898$, and $\sigma_8 = 0.826$~\citep{hinshaw_2013ApJS}.
To construct the convergence maps, the authors use ray tracing under the assumption of flat sky and within the Born approximation~\citep{born2013principles}.
From the convergence maps, the shear maps are determined by leveraging the relationship between $\kappa$, $\gamma_1$, and $\gamma_2$ with the lensing potential function (Equation~\eqref{eq:potential_odes}).
Full detail of the process of going from simulation to maps can be found in~\cite{Harnois-Deraps_2015MNRAS} and~\cite{Harnois-Deraps_2012MNRAS}.

\begin{figure*}[ht!]\label{fig:map_example}
	\centering
	\includegraphics[width=\textwidth]{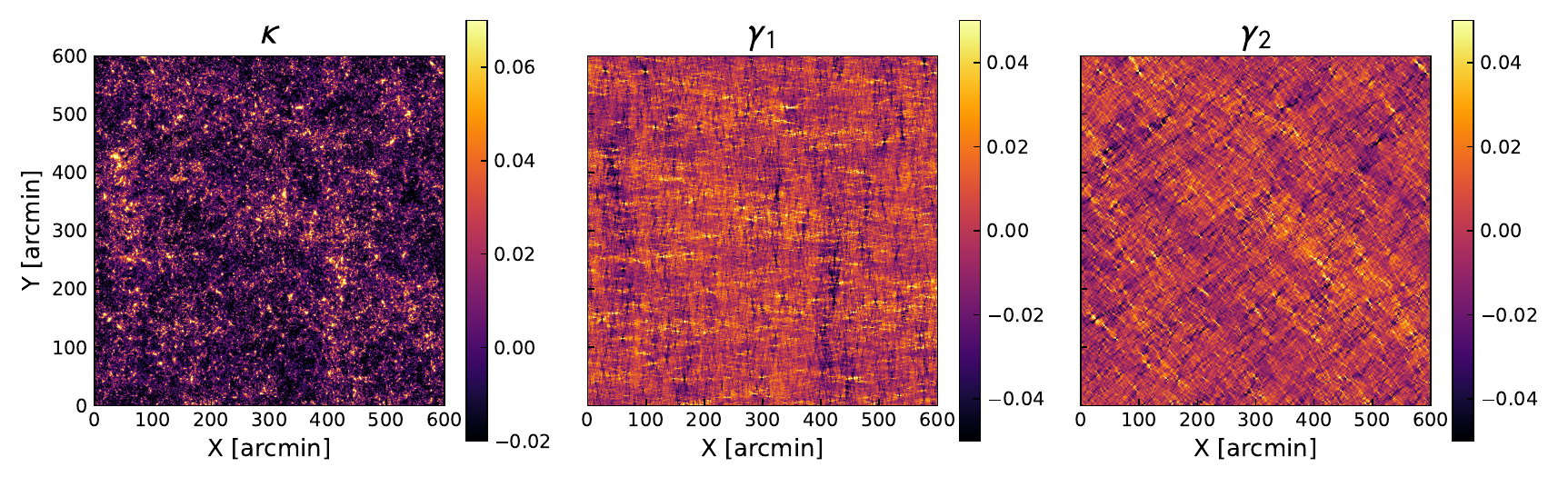}
	\caption{Maps of $\kappa$, $\gamma_1$, and $\gamma_2$ from a single SLICS LOS (LOS420). The maps have been downsampled such that the angular resolution per-pixel is $0.77^2$ square arcminutes and the dynamic range of the image has been altered for clarity.}
\end{figure*}

The data products we use for our analyses include $\sim 1000$ lines of sight (LOS) from SLICS.
A single LOS consists of the information from a $10 \times 10$ degree window of arbitrary sky coordinates from the ray tracing results.
For each LOS, the associated shear and convergence maps, as well as the directly computed matter power spectrum, are included across a range of redshifts, from which we opt to use the information at $z=1.041$.
Shear and convergence maps for a single LOS are of size $7745^2$ pixels, corresponding to a spatial resolution of approximately $0.077^2$ square arcminutes per pixel.
The relevant angular scales covered by the provided shear power spectrum are then on the order of $k \sim [10^1, 10^6]$.

For analysis, we downsample each of the maps to be of size $774^2$ pixels per side, corresponding to a spatial resolution of approximately $0.77^2$ square arcminutes per pixel.
The motivation for this downsampling is two-fold.
First, sub-arcminute length scales are well into the nonlinear regime, which can be hard to model, as well as are not intended to be covered within our model framework, which makes assumptions of linearity in the lensing equations (\ref{sec:method}).
Second, by downsampling the raw SLICS maps before analysis, we are working with $\sim 10^2$ fewer points, which reduces some of the computations needed when scaling to larger datasets.
An example of the $\kappa$, $\gamma_1$, and $\gamma_2$ maps for LOS420 are shown in Figure~\ref{fig:map_example}.

\subsection{Experimental Framework}\label{sec:validate_tests}

Before we test our method at-scale on datasets on the order $\sim 10^9$, we first focus on validating its accuracy at scales on the order $10^6$.
We consider model performance in two primary areas as our metrics for success:
\begin{enumerate}
	\item Shear map residuals,
	\item Recovery of the matter power spectrum as compared to SLICS products.
\end{enumerate}

Shear and convergence maps from SLICS are presented on a regular, finite grid (as shown in Figure~\ref{fig:map_example}) for which we define a set of arbitrary on-sky coordinates.
This entails creating a regular grid of $(x, y)$ coordinates to align with the angular size $\theta$ and granularity of the shear grids.
As these coordinates are completely arbitrary, we define them from $[0,\theta]$ at intervals $\delta \theta = \theta/\ell_{\rm{grid}}$, where $\ell_{\rm{grid}}$ refers to the length of a single side of the grid in pixels.
In order to simulate data from a real survey, we sample ``galaxies'' from the SLICS maps on which we condition the GP.
Galaxy positions are sampled at random spatial points within the window, allowing for samples to be at non-discrete grid positions.
For each sampled $(x, y)$, we compute the corresponding $(\kappa, \gamma_1, \gamma_2)$ with a linear interpolant.

We sample a number of points $N_{\rm{gal}}$ such that it corresponds to our desired galay densities $n_{\rm{gal}}$ such that $n_{\rm{gal}} = N_{\rm{gal}}/\theta^2$ in our flat-sky approximation.
In most cases herein, we will use $n_{\rm{gal}}$ to describe sample sizes, as it is unambiguous across different window sizes.
We test two values of $n_{\rm{gal}}$ from the expectation of LSST at $[5, 30]$ galaxies per square arcminute~\citep{Chang_2013MNRAS}, which correspond to a conservative ``per-redshift-bin'' and whole survey estimate, respectively.

We also test the effect of shape noise on our ability to interpolate accurate maps maps and recover relevant statistics.
We assume a gaussian shape noise of the form $\mathcal{N}(0, \sigma_g)$ where we consider a low-noise regime ($\sigma_g=0.0001$) and a realistic-noise regime ($\sigma_g=0.13$).\footnote{In our framework, we use shear $\gamma_1$, $\gamma_2$ and define our shape noise accordingly as $\sigma_g$. In the case of reduced shear, $g = \frac{\gamma}{1-|\kappa|}$, the shape noise is traditionally interpreted in the context of ellipticity, $\sigma_e$. The mapping between these two noise values is straightforward: $\sigma_e=2\sigma_g$.}
The low-noise regime informs our ability to predict maps accurately at base level while the realistic-noise regime allows us to test the effects of how the density of a galaxy sample affects our accuracy, which will matter when we test our sampled data.

Before performing the GP operations on the input shear maps and spatial grid, we require some extra setup on our inputs.
Our raw inputs are structured as $[\kappa, \gamma_1, \gamma_2]$ across the first tensor dimension, but to adhere to conventions in the operations, we use a $(+, -, +)$ signature, such that we have $[\kappa, -\gamma_1, \gamma_2]$ as our actual inputs to the GP.
We normalize our coordinates to unity such that our galaxy positions are in the range $[0,1]$. For our prediction points, we use the same points as the downsampled map in order to account for issues due to sampling.

It is also important to note how we set up the GP itself, namely the nugget noise and our choice of hyperparameters.
We define our nugget to be homoscedastic with a value defined as the variance of our shape noise $\sigma_g^2$ for $\gamma_1$ and $\gamma_2$ and $\sqrt{2}\sigma_g$ for $\kappa$.
For our hyperparameters, we deviate from a traditional GP study.
At the start of our analysis, we ran tests using a Bayesian optimization method to determine hyperparameter values.
What we found was that the predictions were in fact largely insensitive to the exact value of the $\ell$ hyperparameter~\ref{eq:kernel_psi}.
Within approximately an order of magnitude of a proper value, the resulting maps remain largely unchanged.

As a result, we provide a physical interpretation of the length scale and thus leave it fixed.
In the GP regime, the length scale, from here out denoted $\ell_{\rm{GP}}$, is considered a smoothness parameter.
When generating maps of cosmic shear, it is typical to define a smoothing scale $\ell_{\rm{smooth}}$, which would be the smallest angular scale from which we want to retrieve information and can make conclusions.
There is a mapping from the true physical scale to the GP scale via the normalization: through this we can define our physical smoothing scale and then use the corresponding length in our GP operations.
Note that when we predict onto our test grid, we will have the case that $\delta \theta < \ell_{\rm{smooth}}$, but we cannot make accurate conclusions below those scales.

With respect to the second kernel hyperparameter $\lambda$, which sets the scale of the posterior variance, we leave it fixed as well.
We determine the fixed value via optimization by-hand.

\subsection{Billion-Input Scaling Framework}\label{sec:scale_tests}

As one of the important points of this work, as well as \muygpys{} as a tool, it is important not only that our results be fast
and accurate, but also scalable to large ($\sim10^9$ samples) datasets.
To perform our analysis at scale, we leverage the
Dane\footnote{https://hpc.llnl.gov/hardware/compute-platforms/dane} high performance computing resource along with more of the available SLICS data.
We use a set of one hundred SLICS LOS to create a larger, tiled window of 10,000 square degrees ``on-sky.''
We define our coordinate grid in a manner consistent with Section~\ref{sec:validate_tests}. 
From each LOS we simulate 10 million galaxies randomly
positioned as inputs and predict the shear values at $\sim 600,000$ points. Over the 100 separate LOS, this results in 1
billion simulated inputs and $\sim 60$ million prediction points.  We add Gaussian noise consistent with $\sigma_g = 0.13$.

There is some care needed in distributing the work to different cpus in our workflow.
The number of neighbors can significantly
impact the memory requirements and how many processors per node that we can effectively utilize.
We found a good balance
on the Dane system for 25 nearest neighbors using 56
cores per node on 32 nodes for the nearest neighbor step and 28 cores per node for the prediction step. 
The most demanding step of this is the construction of the nearest neighbors index which takes $\sim 3$ hours in our scale test.

\section{Results}\label{sec:results}

As described in Section~\ref{sec:validate_tests}, we first test the accuracy of our GP predictions at the scale of a single LOS, followed by an extemsion to a larger window consisting of a $10 \times 10$ grid of SLICS LOS from which we sample $10^9$ points.
We present the results at two values of shape noise: $\sigma_g = [0.0001, 0.13]$ across two galaxy sample number densities of $n_{\rm{gal}} = [5, 30] \; \rm{arcmin}^{-2}$.
We assume a smoothing scale $\ell_{\rm{smooth}}=6$ arcminutes and a variance scale of $\lambda=0.125$ for our GP hyperparameters in all analyses.

\subsection{Method Validation}\label{sec:validation}

\begin{figure*}[ht!]
	\centering
	\includegraphics[width=\textwidth]{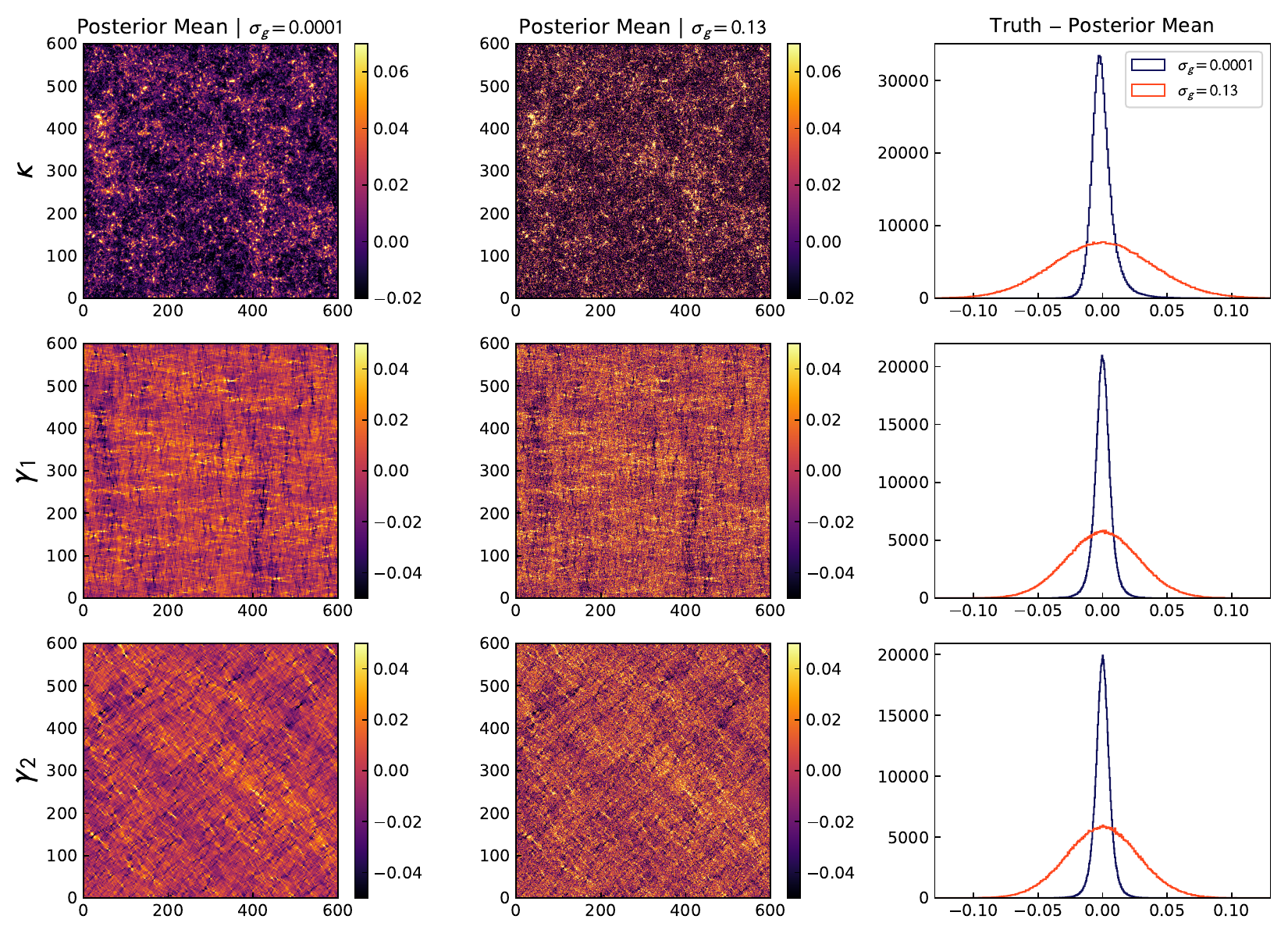}
	\caption{Posterior mean predicted maps for $\kappa$, $\gamma_1$, and $\gamma_2$. Both sets of maps are for $n_{\rm{gal}}=30 \; \rm{arcmin}^{-2}$ and are from the SLICS LOS420. The rightmost column features the histograms of map residuals.\label{fig:nine_pan_results}}
\end{figure*}

To first order, we compare the predicted maps of $\kappa$, $\gamma_1$, and $\gamma_2$ for a set of LOS data.
To compare beyond qualitatively identifying features, we take the difference between the true SLICS map and our posterior mean predictions and look at the distribution of those residuals.
On average, the residuals are centered at zero with a standard deviation that varies based on the $(n_{\rm{gal}}, \sigma_g)$ combination, consistent with our expectations in the given noise regimes.
Comparison of our posterior mean predictions and map residuals for LOS420 are shown in Figure~\ref{fig:nine_pan_results}

Beyond a pure one-to-one map comparison, we are also interested in determining if we preserve the shear correlations to second order.
Namely, we want to compare the shear power spectrum from our convergence map predictions to the data products from SLICS.
To produce the power spectrum, we take the two-dimensional fast Fourier transform of our convergence map and calculate the power in 13 evenly-spaced logarithmic bins in $k$, used here in lieu of $\ell$, as we are using a flat-sky approximation.
In our results, we refer to the \emph{dimensionless} power spectrum, $\Delta_k$, which relates to the power spectrum $P_k$ as described by the Fourier Transform of the two-point correlation function $C_k$ via
\[ \Delta_k = \frac{k^2P_k}{2\pi}. \]
Our choice to use the dimensioness power spectrum is intended to align with the product included in SLICS, which is also the dimensionless representation.
For a complete derivation and proper treatment of these terms, we refer to~\citet{schneider2006introduction}.
We also assume a pure E-mode in these calculations.

\begin{figure*}[ht!]
	\centering
	\includegraphics[width=\textwidth]{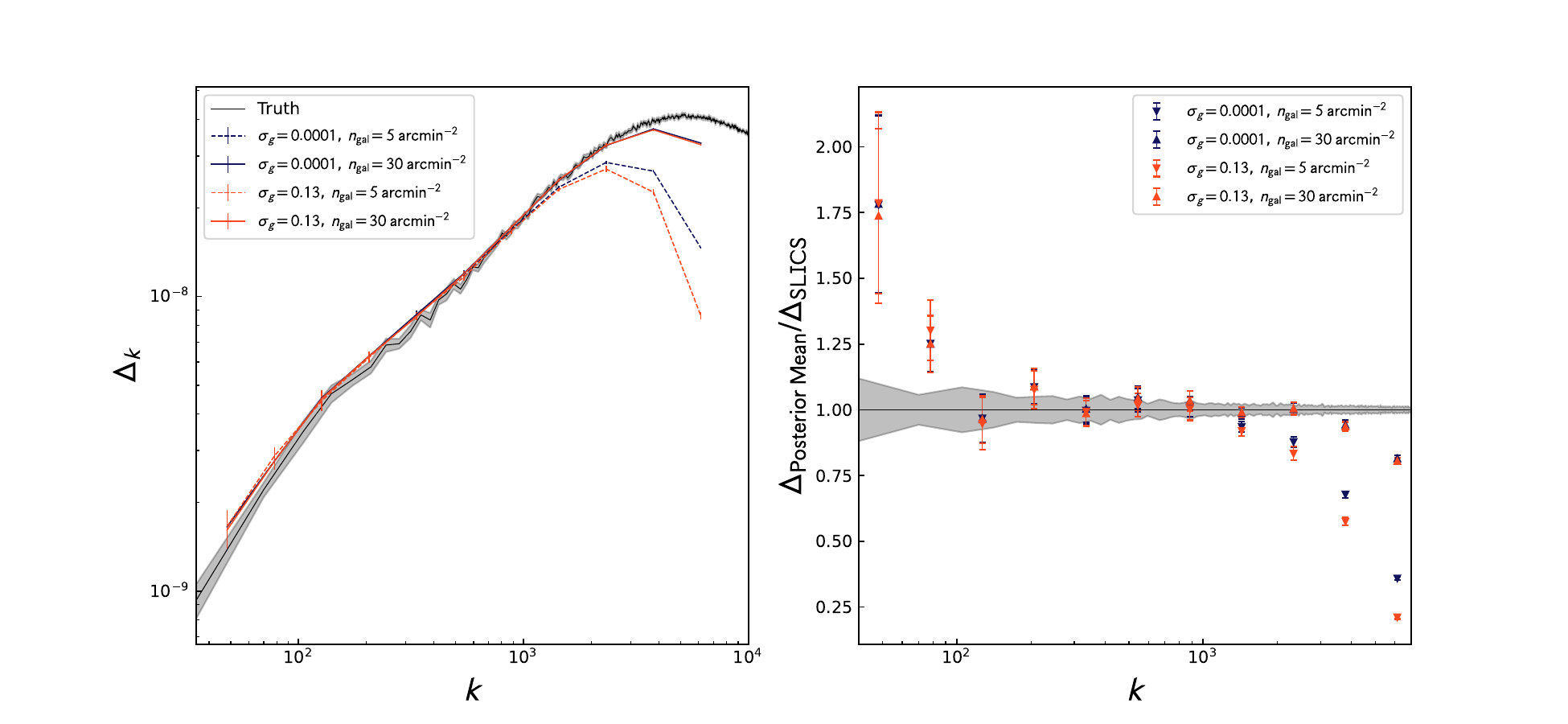}
	\caption{\emph{Left}: Power spectrum results averaged over 21 LOS. The truth values are also averaged over SLICS realizations. \emph{Right}: Ratio of averaged posterior mean predicted $\Delta_k$ to the averaged true power spectrum from SLICS.\label{fig:power_spectrum}}
\end{figure*}

We analyse 21 windows (from LOS400 to LOS420) for each $(n_{\rm{gal}}, \sigma_g)$ configuration.
For each realization, we apply a shot noise correction by taking the power spectrum of a shuffled (uncorrelated) resampling of the window and subtracting it from our signal.
Figure~\ref{fig:power_spectrum} plots our power spectrum results averaged across realizations.
We find that independent of the $(n_{\rm{gal}}, \sigma_g)$ configuration, the power spectrum from our posterior mean predictions agrees with the values calculated directly from the N-body results to within $2\sigma$.
Note that we do not expect to remain consistent with the power spectrum at sub-arcminute scales due to finite sampling effects.
Uncertainty on our estimates is dominated by the inter-realization variance.

\subsection{Billion-Input Scale}

To validate the accuracy of our full scale maps, we focus on the two-point and three-point correlation function. The three-point
function is included to test whether higher order correlations are preserved in our maps.
We use the software package \treecorr\footnote{https://github.com/rmjarvis/TreeCorr}\citep{jarvis2015} for both of these calculations. We
 measure the 2-pt correlation function in 15 logarithmic bins from 1 to 35 arcminutes. For the 3-point function we limit our results to
 equilateral triangles with side length up to $\sim 30$ arcminutes.

\begin{figure*}
	\centering
	\includegraphics[width=0.9\textwidth]{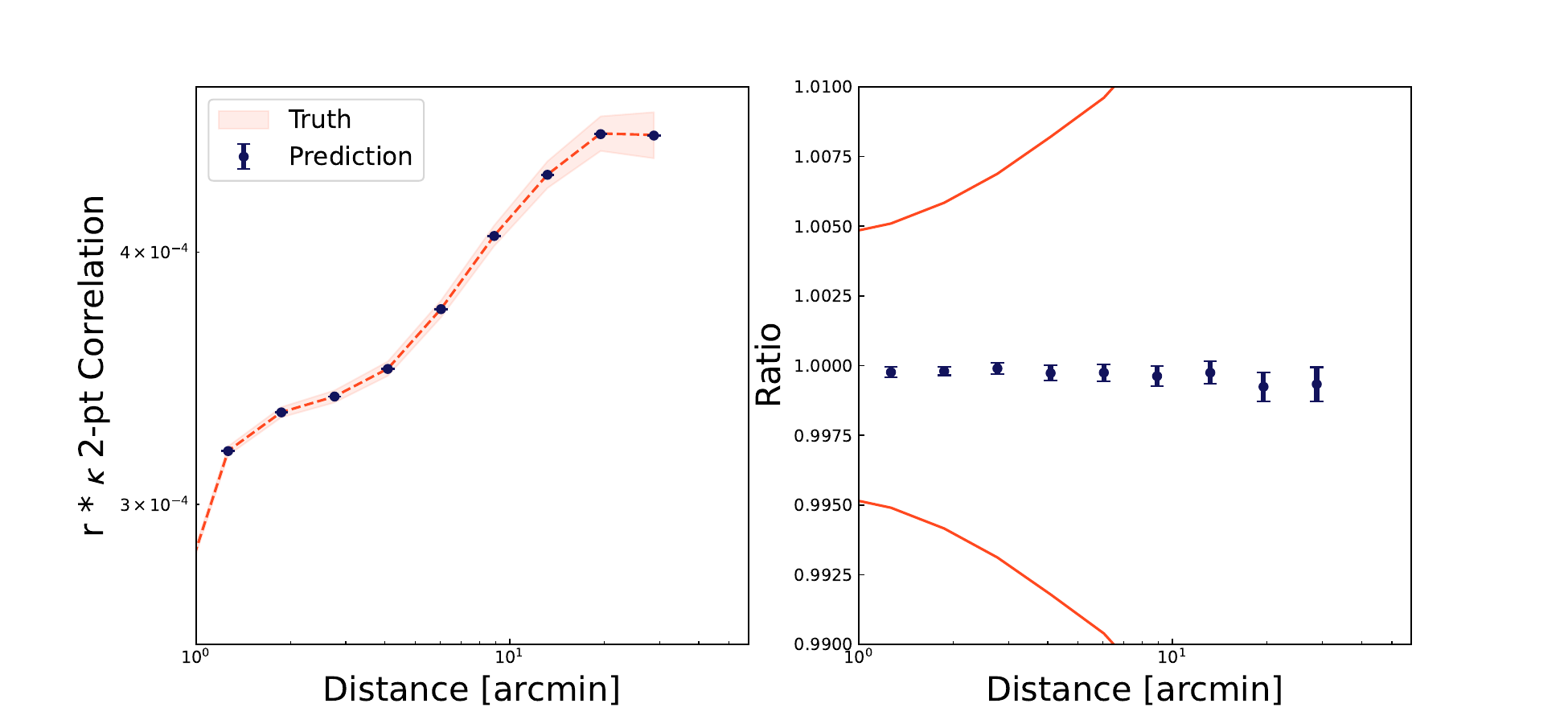}
	\caption{{\it Left}: the 2-point correlation function of the convergence as a function of distance as measured on the full
		10,000$^2$ degree dataset. The red line shows the values as measured by the downsampled true values. The
		 error bars are computed from the variance between the 100 individual LOS. The black line is computed from the
		 posterior variance of our \muygpys code. The error bars on the predictions are computed from resampling from
		 the posterior mean and variance. {\it Right}: the ratio of our predictions to the truth. The red line corresponds to
		 the error on the truth.}
	\label{fig:scale_2pt}
\end{figure*}

We compare our predictions to the true convergence map sampled at the same resolution as our prediction grid.
Figure~\ref{fig:scale_2pt} compares a prediction of the 2-pt correlation function. The predicted 2-pt correlation function
is computed averaging over multiple realizations where we have sampled from the posterior mean and variance for each prediction point.
We also computed the expected uncertainty from the true convergence values by averaging over all the LOS's and scaling the variance appropriately.
This shows that we are in agreement to less than $0.1\%$ on the computed 2-pt correlation function, well below the expected intrinsic uncertainty
from cosmic variance.

\begin{figure*}
	\centering
	\includegraphics[width=0.9\textwidth]{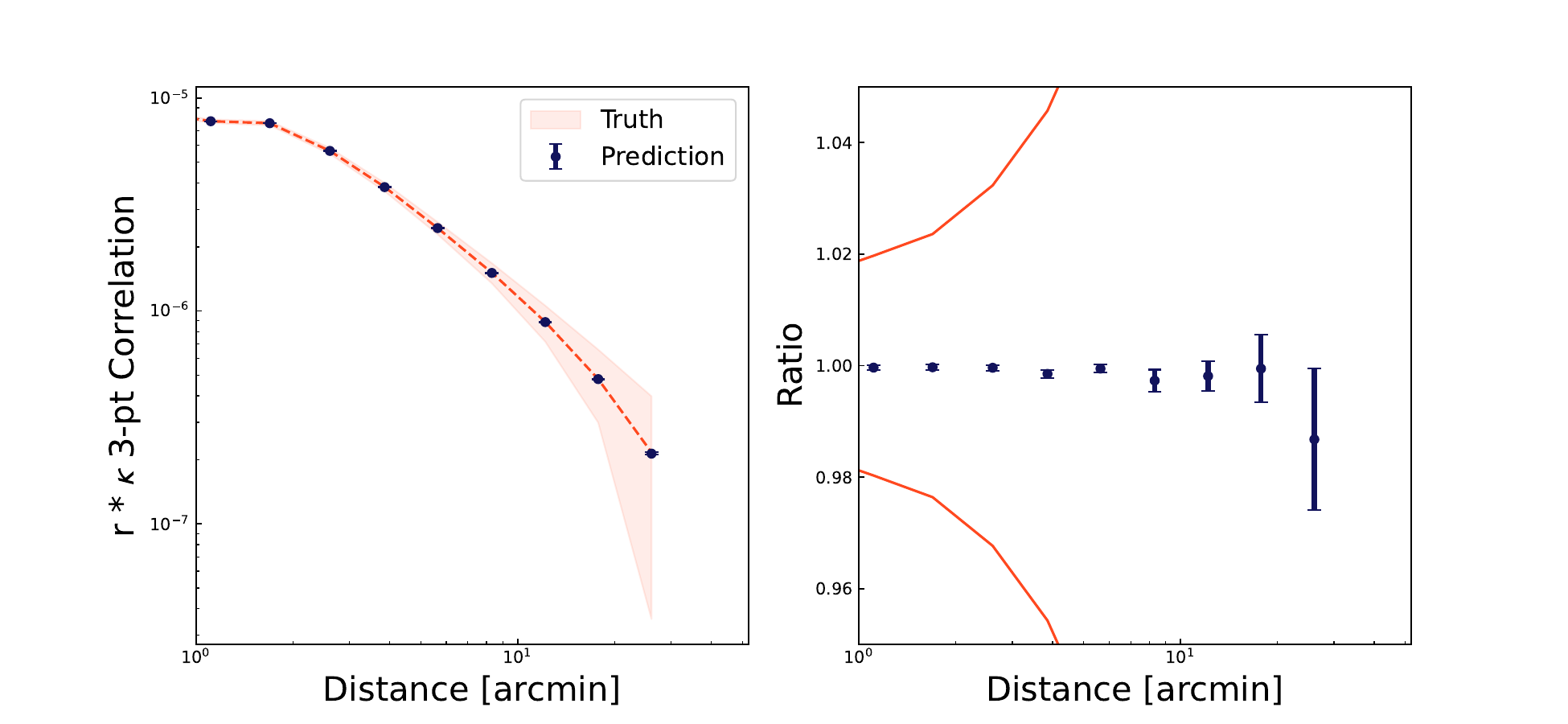}
	\caption{{\it Left}: the 3-point correlation function of the convergence as a function of distance as measured on the full
	10,000$^2$ degree dataset. The triangles are restricted to equilateral triangles. The red line shows the values as measured
	by the downsampled true values. The error bars are computed from the variance between the 100 individual LOS. The black line
	is computed from the posterior variance of our \muygpys code. The error bars on the predictions are computed from resampling
	from the posterior mean and variance. {\it Right}: the ratio of our predictions to the truth. The red line corresponds to
	the error on the truth.}
	\label{fig:scale_3pt}
\end{figure*}

Figure~\ref{fig:scale_3pt} shows a comparison of the measured 3-pt correlation function. As with the 2-pt function,
we average over multiple catalogs generated by sampling from the posterior mean and variance. We see that our prediction of the 3-pt function
is in good agreement with the truth values to a level of less than $1\%$. These results give us confidence that our map predictions maintain the
higher order correlations that can be used to constrain cosmology.

\section{Summary}\label{sec:summary}
We have presented a method of interpolating shear maps using a specialized kernel function within the scalable Gaussian Process library \muygpys{}.
Across our range of noises and galaxy densities, our method is able to accurately recreate the shear maps with a residual distribution consistent with mean zero and a spread $\sim \sigma_g$.

We also scale our training sample to $\sim 10^9$ points.
We find that, at these scales, our results for the two- and three- point correlation functions from our posterior mean map predictions agree with the calculations from SLICS N-body ray-tracing results to within $1\sigma$.
This accuracy in the higher order correlations with the Gaussian Process prior is somewhat unexpected - capturing the three-point function is generally assumed to be in the realm of non-gaussianity in the cosmic matter field.
These suggest that our method will able to lead to informed constraints on cosmology when applied to upcoming survey data.

It is also important to re-emphasize the scale at which these results have been determined: we successfully capture map information at the scale of \emph{one billion} inputs.
Implementing a conventional GP at this scale is entirely impractical, but in the MuyGPs framework it is, as we have demonstrated, wholly achievable.

\section{Future Work}\label{sec:future}

The results presented in this work are contingent on having data from both shear and convergence on which to condition the GP.
However, in a survey environment, it is seldom the case to have knowledge of the convergence \textit{a priori} (although several studies have proposed using magnification to get information on convergence, for example \cite{huff2014, alsing2015, bernstein2016}).
A natural extension of this method for large surveys without convergences is to formulate the kernel such that we condition the GP on only $\gamma_1$ and $\gamma_2$ in order to predict the map of convergence.
Preliminary testing suggests that at least a single measure of convergence is necessary to set the proper scale for the convergence predictions. We will explore this in more detail in future work.

Additionally, in weak lensing under the Born Approximation~\citep{born2013principles}, we generally assume the shear fields are curl-free (a pure E-mode), but in practice there are various effects (both systematic and astrophysical) that can introduce a B-mode.
For example, any intrinsic alignment of the source galaxies~\citep{troxel_2015PhR}, higher-order terms and relations not present in the linearized lensing regime, and systematics in image analysis can all induce a B-mode.

In the effort of disentangling the E- and B-mode power spectra, the lensing poential can be decomposed into a real and complex part, such that $\Psi(\mathbf{\theta}) = \psi^E(\mathbf{\theta}) + i\psi^B(\mathbf{\theta})$, where $\psi^E$ and $\psi^B$ refer to the E- and B-mode contributions, respectively.
In our framework, this analysis would entail recasting the kernel function to include this complex component.
Separation of the E- and B-mode components allows for more pointed investigation of the systematic and astrophysicsal sources in the data which could be contributing to B-mode contamination.

In our analysis, we have also worked entirely under the assumption that the sky coordinates can be accurately described my a Minkowski metric.
While this assumption holds for small enough scales, extension to a survey environment of $\gtrsim 100^2$ square degrees requires a more careful treatment of our coordinate system.
We save the derivation and expression of the lensing equations and our kernel function in a curved metric for our follow up work.

While we have presented evidence of the accuracy and validity of our method under the aforementioned assumptions, we strive to create a more robust framework for shear mapping with our GP approach.
With the combination of a convergence-from-shear form of the kernel and the option to work in a curved metric or with complex potential to disentangle E- and B-mode contributions to the power spectrum, our method will be widely applicable to data from next-gen surveys.

\section*{Acknowledgements}
\begin{acknowledgements}
	This work was performed under the auspices of the U.S. Department of Energy by Lawrence Livermore National Laboratory under Contract DE-AC52-07NA27344 (LLNL-JRNL-869702), and was supported by LLNL LDRD project 22-ERD-028.

	The authors would like to thank Joachim Harnois-D{\'e}raps for helping access the SLICS data used in this work.
\end{acknowledgements}

\bibliography{paper}{}
\bibliographystyle{aasjournal}

\end{document}